\documentstyle[twoside,fleqn,espcrc2]{article}

\newcommand{\n}{\hspace*{-2.5mm}}

\newcommand{\AmS}{{\protect\the\textfont2
  A\kern-.1667em\lower.5ex\hbox{M}\kern-.125emS}}

\title{\hfill{\small TUM--HEP--249/96}\\
       \hfill{\small hep-ph/9608356}\\[3mm]
Higher-order corrections in the SM Higgs sector: the right scale
\thanks{To appear in the Proceedings of the Workshop ``QCD and QED in Higher
Order'', Rheinsberg, Germany (April 1996).}}
\author{Kurt Riesselmann
\address{Physik Department, Technische Universit\"at M\"unchen,\\ 
        James-Franck-Str., D-85748 Garching, Germany}%
}

\begin{document}
\thispagestyle{empty}

\begin{abstract}
  The evaluation of high-energy cross sections involving the SM Higgs boson
  requires the use of the Higgs running coupling $\lambda(\mu)$. Taking $\mu$
  to be equal to the center-of-mass energy $\sqrt{s}$ of the scattering
  process, the perturbative approach fails for relatively small values of the
  Higgs mass and coupling, $\lambda(\sqrt{s})\approx 2.2$.  Performing an
  approximate resummation of ``bubble'' Feynman diagrams, we find the scale
  $\mu=\sqrt{s}/{\rm e}\approx \sqrt{s}/2.7$ to yield reliable perturbative
  results, even for large Higgs mass and coupling. The new perturbative upper
  limit on the Higgs running coupling is $\lambda(\sqrt{s}/{\rm e})\approx 4$.
\end{abstract}

\maketitle

\section{Physical motivation}

The Higgs boson of the Standard Model (SM) still eludes detection.  Once a
Higgs boson has been discovered, its couplings to other particles as well as
its self-couplings need to be measured and compared with SM predictions.  Of
particular interest are the couplings of the Higgs boson to the (longitudinally
polarized) gauge bosons $W^+, W^-, Z$ of the SM.  Scattering processes
involving these particles receive contributions from the gauge couplings (which
already can be tested at LEP2) as well as the trilinear and quartic Higgs
couplings.  To test the latter, reliable predictions for SM cross sections and
decay widths are important.  In the case of high-energy scattering processes,
the resummation procedure described here is essential.  The prospects for
studying the various couplings at future colliders are, for example, discussed
in \cite{lep2,lhc,self1,self2,nlc}.

At the same time, physical observables involving Higgs couplings can be used to
derive upper bounds on a perturbative Higgs coupling. The size of the radiative
corrections, the sensitivity of the result to the renormalization scale $\mu$,
and the perturbative violation of unitarity give upper limits on the values of
the Higgs mass $M_H$ and the Higgs quartic coupling $\lambda$ beyond which
perturbation theory fails.  Using high-energy amplitudes, rather stringent
bounds on the Higgs quartic coupling have been
derived \cite{unit1,unit2,rie,nie}.  We show that these bounds are relaxed when
the right choice of scale is chosen in connection with the running coupling.

\section{The Higgs running coupling}

In the SM, the value of the quartic Higgs coupling $\lambda$ is fixed at tree
level by the value of the Higgs mass: $\lambda=M_H^2/2v^2$, where $v=246$ GeV
is the vacuum expectation value.  Calculating physical observables with energy
scales larger than $M_H$, renormalization group methods suggest the use of the
running coupling $\lambda(\mu)$.  Fixing the running coupling at the scale
$\mu=M_H$ such that it equals the tree level result,
\begin{equation}
\lambda(M_H)=\frac{M_H^2}{2v^2}\,, 
\end{equation}
the value of the running coupling at scales $\mu>M_H$ is fixed by the $\beta$
function of the theory.  Neglecting gauge and Yukawa couplings, a good
approximation for a heavy Higgs boson, the one-loop result for the running
coupling is
\begin{equation}
\lambda(\mu)=\frac{\lambda(M_H)}
{1 - \frac{\beta_0}{2}\,\frac{\lambda(M_H)}{16\pi^2}
\,\ln\!\left(\frac{\mu^2}{M_H^2}\right)}
\,,\quad  \beta_0=24\,.
\label{runcoup}
\end{equation}
Solutions up to three loops are discussed in \cite{nie}.  From the previous
equations it is apparent that perturbation theory will cease to be useful if
either the Higgs mass $M_H$ is large or if the energy scale $\mu$ of a process
is large.

\section{Two-loop results in the Higgs sector}

Several observables related to the SM Higgs sector have been calculated to two
loops in perturbation theory using the heavy-Higgs limit.  Neglecting the
subleading corrections due to gauge and Yukawa couplings, the two-loop
$O(\lambda^2)$ results for the partial widths of the Higgs boson decaying into
a pair of gauge bosons \cite{hww1,hww2} or a pair of fermions \cite{hff1,hff2}
are:
\begin{eqnarray}
\label{hwwwidth}
\Gamma(H&\n\rightarrow\n& W^+W^-) \propto \nonumber\\
\lambda(M_H)\n\n&\n\n\n&\left(1
+ 2.8 \frac{\lambda(M_H)}{16\pi^2} 
+ 62.1 \frac{\lambda^2(M_H)}{(16\pi^2)^2}\right)\,,\\
\Gamma(H&\n\rightarrow\n& f\bar f) \propto \nonumber\\
g_f^2 \n\n&\n\n& \left(1
+ 2.1\frac{\lambda(M_H)}{16\pi^2}
- 32.7 \frac{\lambda^2(M_H)}{(16\pi^2)^2}\right)\,.
\end{eqnarray}
Here $g_f$ is the Yukawa coupling of the fermion $f$.  Comparing the magnitude
of the one-loop and two-loop corrections, perturbation theory seems to work up
to values of $\lambda(M_H)\approx 7$, that is, Higgs masses of about 1 TeV.
However, an analysis of the scale- and scheme-dependence \cite{nie} reveals
that higher-order terms may spoil perturbation theory already for values of
$\lambda(M_H)\approx 4$, or equivalently, $M_H\approx 700$ GeV.

Perturbation theory fails even sooner in the case of high-energy scattering
processes.  For example, the two-loop correction to the scattering process
$W_L^+W_L^-\rightarrow Z_LZ_L$ is \cite{rie,pnm}
\begin{eqnarray}
\sigma(W_L^+W_L^-&\n\rightarrow\n& Z_LZ_L)\propto \nonumber\\
\lambda^2(\sqrt{s})\!&\n\n&\n\n\left(1 - 42.6
\frac{\lambda(\sqrt{s})}{16\pi^2} 
+ 2457.9 \frac{\lambda^2(\sqrt{s})}{(16\pi^2)^2}\right)
\label{cross}
\end{eqnarray}
where the subscript $L$ denotes the longitudinal polarization of the gauge
bosons, and $\sqrt{s}$ is the center-of-mass energy of the incoming particles.
The assumption $\sqrt{s}\gg M_H$ used in calculating (\ref{cross}) is valid if
$\sqrt{s}>2-3 M_H$ \cite{rie}. It is striking that the coefficients appearing
in the cross section are much larger than the corresponding coefficients of the
decay widths.  Consequently, the upper bound on a perturbative Higgs coupling
is stronger than in the case of the decay widths.  Comparing the size of the
one-loop and two-loop cross section, a bound of $\lambda(\sqrt{s})\approx 2.2$
is found \cite{rie}. A similar bound is also derived when investigating the
scheme- and scale-dependence of the cross section \cite{nie}. Since
$\lambda(\sqrt{s}) > \lambda(M_H)$ for $\sqrt{s} > M_H$, the bound
$\lambda(\sqrt{s})\approx 2.2$ represents a strong restriction on the Higgs
mass.  Choosing $\sqrt{s}$ to be a couple of TeV, the Higgs mass has to be less
than about 400 GeV to allow for a perturbative calculation of the cross
section.
 
The cross sections for low energies ($\sqrt{s}< M_H$) and energies near the
resonance ($\sqrt{s}\approx M_H$) have been calculated including one-loop
corrections \cite{low1,low2,low3}. The calculation requires the inclusion of
trilinear Higgs coupling.  The one-loop results do not yield stringent bounds
on the Higgs coupling. A two-loop calculation is not yet available.

\section{Resummation of bubble contributions: the right scale for scattering
processes }

We are able to improve the perturbative character of the cross section
(\ref{cross}) significantly by introducing a resummation of nonlogarithmic
terms.

\subsection{Identification of bubble contributions}

The large size of the one- and two-loop coefficients in (\ref{cross}) can be
traced back to the finite pieces of the scattering graphs which contribute to
the amplitude. In the limit $\sqrt{s}\gg M_H$, {\it all} one-loop scattering
graphs can be written in terms of the {\it massless} bubble Feynman diagram
$B(p^2)$, the one-loop two-point function with two massless propagators. The
quantity $p^2$ is the four-momentum squared running through the bubble.  Using
the usual Mandelstam variables $s,t,u$, the high-energy amplitude of the
scattering process considered above can be written as
\begin{eqnarray}
a\!&\n(\n&\!W_L^+W_L^-\rightarrow Z_LZ_L) =\nonumber\\
&\n-\n&\! \lambda(\mu)\left[ 2 + O(M_H^2/s)\right]\nonumber\\
&\n-\n&\! \lambda^2(\mu)
\left[ 16B(s) + 4B(t) + 4B(u) + O(M_H^2/s)\right]\nonumber\\
&\n+\n& {\rm counterterms} +\; {\rm wavefct. renorm.}\,\,.
\label{ampstrt}
\end{eqnarray}
The counterterms and wavefunction renormalizations are independent of $s,t,u$.
The massless one-loop bubble is evaluated in dimensional regularization.
Choosing the dimension to be $D=4-2\epsilon$, the expression 
for $B(p^2)$ equals
\begin{eqnarray}
\!\n&\n B\n&\!(p^2) =\nonumber\\
\!\!\!&\n\n\n&\n\! \frac{(4\pi e^{-\gamma})^\epsilon}{16\pi^2} 
\!\left(  \frac{1}{\epsilon} + \ln\frac{\mu^2}{|p^2|} + 2 
+i\pi\Theta(p^2)
+ O(\epsilon)\right)\!\!.
\label{bubble}
\end{eqnarray}
The step function $\Theta$ ensures that only the $s$-channel gives rise to an
imaginary part.

The variables $t$ and $u$ are proportional to $s$ and contribute to the
logarithmic $s$-dependence of the amplitude due to the bubble diagrams.  The
coeffcients of all $s$-dependent logarithms add up to the one-loop beta
function coefficient $\beta_0=24$.  The complete resummation of these leading
logarithms $\ln(s)$ is achieved by setting the scale $\mu$ of the running
coupling to be equal to $\sqrt{s}$, the standard choice in renormalization
group procedures.  Nonlogarithmic pieces do not get resummed when making this
choice.
 
\begin{figure}[b]
\vspace{2.05cm}
\includegraphics{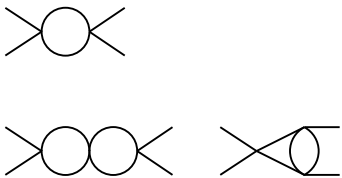}
\vspace{0.7cm}
\caption{
  Topologies of $s$-channel Feynman diagrams contributing to high-energy
  $W_L^+W_L^-\rightarrow Z_LZ_L$ scattering at one loop (top row) and two loops
  (bottom row).  }
\label{figdiag}
\end{figure}

Looking at (\ref{bubble}) we find that the logarithm is accompanied by a
constant 2.  While the imaginary part of the bubble only contributes to the
$s$-channel, the constant 2 is universal to $s$, $t$, and $u$ channel. It
contributes to the one-loop amplitude a term $2\beta_0=48$. Including
counterterms and wavefunction renormalizations, the complete high-energy
amplitude is
\begin{eqnarray}
a= -2\lambda(\mu) + \frac{\lambda^2(\mu)}{16\pi^2}&\n\n&\!\left[\,
\beta_0\left(\ln\frac{\mu^2}{s} + 2\right) + 16i\pi\right. \nonumber\\
&\n\n& \left.- 4\ln\frac{\sin^2\theta}{4} - 13.35\, \right]\,,
\label{ngamp}
\end{eqnarray}
where $\theta$ denotes the c.m.s. scattering angle which relates $t$ and $u$ to
$s$.  We find that the contribution from the 2's dominates the total one-loop
correction of the amplitude and the cross section.  It is important to note
that {\it each} logarithm of the one-loop amplitude is accompanied by the
constant 2 and that there is no further contribution proportional to $\beta_0$.
In particular, the counterterms and wavefunction renormalizations in
(\ref{ampstrt}) are evaluated from a variety of different low-energy Feynman
diagrams, all of which have different nonlogarithmic pieces.  Details are given
in \cite{wil}.

At higher orders in perturbation theory, the ``massless bubble structure'' of
the high-energy scattering graphs persists.  This is due to the $\Phi^4$-theory
nature of the high-energy interactions in the SM Higgs sector.  In
Fig.~\ref{figdiag} we see the $s$-channel Feynman diagrams of the $2\rightarrow
2$ amplitude at one and two loops. The $t$- and $u$-channel contributions are
not shown. They are obtained using crossing symmetries of the external legs.
At one loop, the bubble is the only scattering topology, yielding the
contributions according to (\ref{ampstrt}).  At two loops, we encounter two
topologies: the squared bubble topology $B^2(p^2)$ and the insertion of a
bubble at the vertex of a second bubble.  In the case of the topology $B^2$,
each bubble gives rise to a constant 2.  In the second topology, only the inner
bubble features a constant 2. The integration of the outer loop is not a simple
bubble integral.  It yields a finite piece different from 2.

The situation is similar at three loops and beyond.  At $n$ loops there always
is a topology which is a product of $n$ bubbles.  Next there are topologies
which have $n-1$ bubbles, with the final integration being a modified bubble
integral.  Then there are topologies with $n-2$ bubbles, and so on.  Starting
at three loops, there also exist nonplanar graphs which cannot be naturally
viewed as being constructed from bubble graphs.  Yet their weight is expected
to be small compared to the numerous bubble related contributions to the
complete set of $n$-loop diagrams.

\subsection{The right scale}

The above arguments suggest that the constant 2 accompanies the majority of the
leading logarithms. The resummation of this contribution is desirable.  Since
the $s$-dependend leading logarithms are resummed using the choice
$\mu=\sqrt{s}$, it is easy to adjust $\mu$ to also yield a resummation of the
``leading 2'' such that
\begin{equation}
\ln\!\left(\frac{\mu^2}{s}\right) + 2 = 0\,.
\end{equation}
The right scale choice is thus
\begin{equation}
\mu=\frac{\sqrt{s}}{\rm e}\approx\frac{\sqrt{s}}{2.7}\,.
\end{equation}
The resulting leading-log (LL) result for the high-energy amplitude is
therefore
\begin{equation}
a(W_L^+W_L^-\rightarrow Z_LZ_L)= -2\lambda(\sqrt{s}/{\rm e})\,.
\end{equation}

\subsection{Testing the new choice of scale}

Such a particular choice of $\mu$ is, of course, only meaningful if one expects
it to work at higher orders as well.  An exact resummation of subleading
logarithms together with nonlogarithmic terms is not possible. All known
scale-setting schemes resort to various methods to justify their choice of
scale at subleading level; see \cite{sch} for a discussion.  A first check for
the validity of the leading-log scale choice at higher orders is the size of
the coefficients at higher orders. A bad choice of scale would reduce the
perturbative coefficients in lower orders and push large corrections into
higher orders. Taking our scale choice $\mu=\sqrt{s}/{\rm e}$, we re-evaluate
the next-to-next-to-leading-log (NNLL) cross section (\ref{cross}):
\begin{eqnarray}
\label{crossimprov}
\sigma(W_L^+W_L^-&\n\rightarrow\n& Z_LZ_L)\propto \\
\n\n\n\lambda^2(\sqrt{s}/{\rm e})&\n\n&\n\n\left(1 + 5.4
\frac{\lambda(\sqrt{s}/{\rm e})}{16\pi^2} 
+ 539.9 \frac{\lambda^2(\sqrt{s}/{\rm e})}{(16\pi^2)^2}\right)\,.\nonumber
\end{eqnarray}
Comparing the coefficients with the result for $\mu=\sqrt{s}$ in (\ref{cross})
we find that both one-loop and two-loop coefficient are greatly reduced in
magnitude.  This suggests that our improved scale choice also works at the
subleading level.

\begin{figure}[t]
\vspace{6.8cm}
\includegraphics{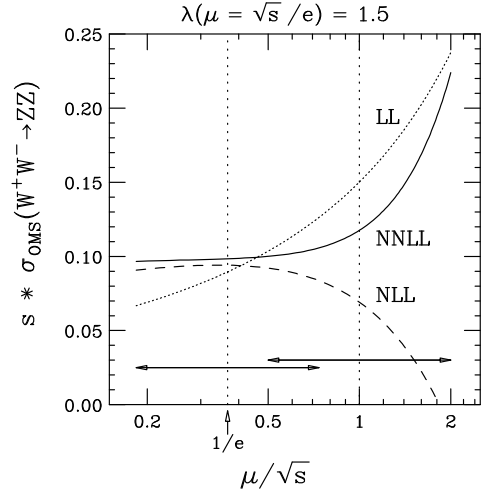}
\caption{
  The scaled high-energy cross section of $W_L^+W_L^-\rightarrow Z_LZ_L$ for
  $\lambda=1.5$.  Gauge and Yukawa coupling contributions are neglected.  The
  horizontal lines with arrows indicate the typical variation of $\mu$, taking
  the central values to be $\mu=\protect\sqrt{s}/{\rm e}$ and
  $\mu=\protect\sqrt{s}$.  }
\label{figcross}
\end{figure}

Another good check on the quality of our scale choice is the variation of the
scale around the central value chosen. The explicit $\mu$-dependence of the
cross section $\sigma(W_L^+W_L^-\rightarrow Z_LZ_L)$ can be found in
\cite{nie}.  In Fig.~\ref{figcross} we show the LL, NLL, and NNLL results for
$s\,\sigma$ as a function of $\mu/\sqrt{s}$. The Higgs running coupling is
fixed to be $\lambda(\mu=\sqrt{s}/{\rm e})=1.5$. ( The coupling at
$\mu=\sqrt{s}$ is then fixed by Eq.~(\ref{runcoup}) and yields
$\lambda(\sqrt{s})=1.9$.) Such a value corresponds, for example, to a Higgs
mass of 390 GeV and $\sqrt{s}=$1 TeV.  Fixing the running coupling at a certain
scale, the quantity $s\,\sigma$ is a function of the single variable
$\mu/\sqrt{s}$.  Varying $\mu$ around the central values $\sqrt{s}$ and
$\sqrt{s}/{\rm e}$ as indicated in Fig.~\ref{figcross}, we determine the
sensitivity of the result with respect to the renormalization group logarithms.
If perturbation theory is to be reliable, one expects an order-by-order
reduction of the scale dependence.  Comparing the results of our scale choice
$\mu=\sqrt{s}/{\rm e}$ with the standard choice $\mu=\sqrt{s}$ we find a
greatly reduced scale dependence when choosing $\sqrt{s}/{\rm e}$.  The scale
dependence around $\mu=\sqrt{s}/{\rm e}$ nicely reduces when going from LL to
NLL order.  At the same time, the magnitude of the one-loop and two-loop
corrections is significantly reduced when using our improved scale choice, a
fact that we already noticed by comparing (\ref{cross}) and
(\ref{crossimprov}).

Looking at Fig.~\ref{figcross} we also see that the NLL cross section can
become negative, that is, the magnitude of the one-loop correction exceeds
unity if the scale $\mu$ is chosen too large: The perturbative calculation
completely fails.  The choice $\mu=\sqrt{s}/{\rm e}$, however, is clearly in a
perturbatively reliable region.

\subsection{Upper bound on the running coupling}

Of course, the perturbative behaviour of the cross section also depends on the
value of the running coupling. In Fig.~\ref{figcross} we chose
$\lambda(\sqrt{s}/{\rm e})=1.5$.  Increasing this value, perturbation theory
will eventually also cease to be useful even when taking our improved choice of
scale.  We have found that this happens for $\lambda(\sqrt{s}/{\rm e})\approx
4$. Such a value corresponds to, for example, $M_H=700$ GeV and $\sqrt{s}=1.9$
TeV.  (A larger value of $M_H$ results in a smaller value of $\sqrt{s}$,
eventually violating the high-energy assumption of the calculation.) The
details on obtaining the upper bound are given in \cite{wil}, where a
re-analysis of unitarity violation is carried out.  Comparing the new bound
with the previous result of $\lambda(\sqrt{s})\approx 2.2$
\cite{unit1,unit2,rie,nie}, the perturbative region is significantly extended.

There also exist upper bounds on the Higgs mass and coupling which are derived
from the renormalization group equations for the Higgs running coupling.
Requiring that the one-loop running coupling $\lambda(\Lambda)$ remains finite
up to large embedding scales $\Lambda$, an upper bound on $\lambda(M_H)$ is
deduced \cite{rge}.  These bounds are less stringent than the bounds derived
from physical observables such as cross sections and decay widths. 

The constraints obtained from lattice calculations are described in
\cite{lat}. A comparison of perturbative and lattice results is given
in \cite{wil}.

\section{The right scale in Higgs decays}

The necessity of modifying the standard scale choice in scattering processes
leads to the question of whether the standard scale choice in decay processes
needs to be modified as well.  In contrast to scattering processes, the Higgs
decays feature a fixed center-of-mass energy of $\sqrt{s}=M_H$.  The Higgs
mass appearing in the Feynman integrals cannot be neglected.  Consequently,
the different bubble topologies contribute various finite pieces to the
one-loop result: no universal term proportional to $\beta_0$ arises. This
suggests to leave the standard choice $\mu=M_H$ unchanged.

\begin{figure}[t]
\vspace{5.8cm}
\includegraphics{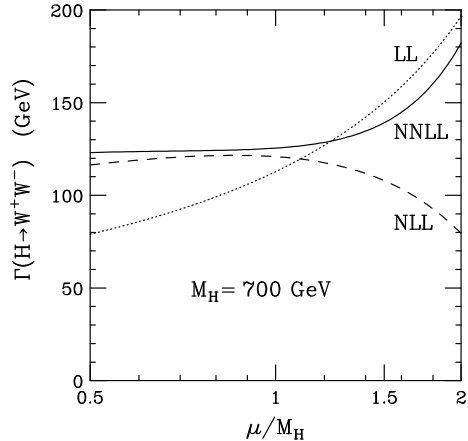}
\caption{
  The decay width of $H\rightarrow W_L^+W_L^-$ varying the scale $\mu$. The
  Higgs mass is fixed to be 700 GeV.}
\label{figwidth}
\end{figure}

Similar to the discussion of the cross section, we can check the correctness of
the choice $\mu=M_H$ by plotting the $\mu$-dependence of the decay widths.  The
$\mu$-dependence of the two-loop width (\ref{hwwwidth}) is given in \cite{nie}.
Fixing the Higgs mass to be 700 GeV, the result for the decay channel
$H\rightarrow W^+W^-$ is given in Fig.~\ref{figwidth}.  The scale $\mu=M_H$
seems to be a good choice.  A major modification like in the case of
high-energy scattering processes is not necessary.

\section{Summary}

If the Higgs mass is large, perturbative high-energy cross sections such as
$\sigma(W^+_LW^-_L\rightarrow Z_LZ_L)$ are sensitive to the choice of the
renormalization scale $\mu$.  The usual choice $\mu=\sqrt{s}$ can lead to
unreliable perturbative results for relatively small Higgs mass and coupling.
Motivated by the bubble structure of the contributing Feynman diagrams, we find
the choice $\mu=\sqrt{s}/{\rm e}\approx\sqrt{s}/2.7$ to yield an approximate
resummation of specific terms in the perturbative expansion.  As a result we
obtain reliable perturbative predictions even for Higgs couplings as large as
$\lambda(\sqrt{s}/{\rm e})\approx 4$, a significant improvement over the
previous bound of $\lambda(\sqrt{s})\approx 2.2$. In the case of Higgs decays
no dominant bubble contributions exist, and the standard choice $\mu=M_H$ is
the right choice.

\vspace{5.5mm} 
{\noindent \bf Acknowledgements} 
\vspace{3.5mm} 

I thank the theory group of DESY-IfH Zeuthen for inviting me to this workshop.
It was a pleasure to introduce the recent work on radiative corrections in the
SM Higgs sector to the experts of the QED and QCD community.  Discussions with
many participants of the workshop are gratefully acknowledged, especially
conversations with K.G.  Chetyrkin and A. Davydychev.  Further thanks go to my
collaborator Scott Willenbrock.

\end{document}